\documentclass[aps,superscriptaddress,twocolumn]{revtex4-1}

\usepackage{mathrsfs}
\usepackage{amsfonts}
\usepackage{amssymb}
\usepackage{amsmath}
\usepackage{graphicx}
\usepackage[usenames,dvipsnames]{color}
\usepackage[colorlinks=true,citecolor=blue,linkcolor=magenta]{hyperref}
\usepackage{ulem}
\usepackage{lmodern}

\newcommand{\Tr}{\mathrm{Tr}}

\newcommand{\abs}[1]{| #1 |}

\newcommand{\mean}[1]{\langle #1 \rangle}

\begin{document}

\title{Interference-based molecular transistors}

\author{Ying Li}

\affiliation{Department of Materials, University of Oxford, Parks Road, Oxford OX1 3PH, United Kingdom}

\author{Jan A. Mol}

\affiliation{Department of Materials, University of Oxford, Parks Road, Oxford OX1 3PH, United Kingdom}

\author{Simon C. Benjamin}

\affiliation{Department of Materials, University of Oxford, Parks Road, Oxford OX1 3PH, United Kingdom}

\author{G. Andrew D. Briggs}

\affiliation{Department of Materials, University of Oxford, Parks Road, Oxford OX1 3PH, United Kingdom}

\date{\today}

\begin{abstract}
Molecular transistors have the potential for switching with lower gate voltages than conventional field-effect transistors. We have calculated the performance of a single-molecule device in which there is interference between electron transport through the highest occupied molecular orbital and the lowest unoccupied molecular orbital of a single molecule. Quantum interference results in a subthreshold slope that is independent of temperature. For realistic parameters the change in gate potential required for a change in source-drain current of two decades is 20 mV, which is a factor of six smaller than the theoretical limit for a metal-oxide-semiconductor field-effect transistor. 
\end{abstract}

\maketitle

\section{Introduction}

No technology is further from its thermodynamic limit than the switching that forms the basis of all information and communication technologies (ICT). A typical CMOS logic operation dissipates about $1 \,{\rm fJ}$, which is 300,000 times the Landauer minimum $k_\text{B}T \ln 2 = 3 \,{\rm zJ}$ at room temperature. In a well designed device the dissipation of energy in a switching operation is dominated by a term associated with capacitive charging which scales as the voltage squared. The operating voltage is in turn constrained by the subthreshold swing, the change in gate voltage $V_\text{g}$ required to change the source-drain current by a factor of ten. This is limited by the exponential tail in the Fermi-Dirac distribution of the thermally excited electrons passing over the barrier created by the gate, which in the best case gives a current $\propto e^{-eV_\text{g}/k_\text{B}T}$~\cite{SzeBook},~i.e. a subthreshold swing of $60 \,{\rm mV}/{\rm decade}$. This thermal limitation can be overcome by exploring quantum effects, as is done in tunnel-field effect transistors~\cite{Appenzeller2004, Seabaugh2010}. In this paper, we provide a theoretical study, with a calculation for specific parameters, to show how quantum interference effects in single-molecule devices~\cite{Reed1997, Park2002, Kubatkin2003, Sedghi2011, Guedon2012, Sotthewes2013} could be used to give vast improvements in the subthreshold swing, and hence in the energy consumption of logic circuits.

\begin{figure}[tbp]
\centering
\includegraphics[width=1\linewidth]{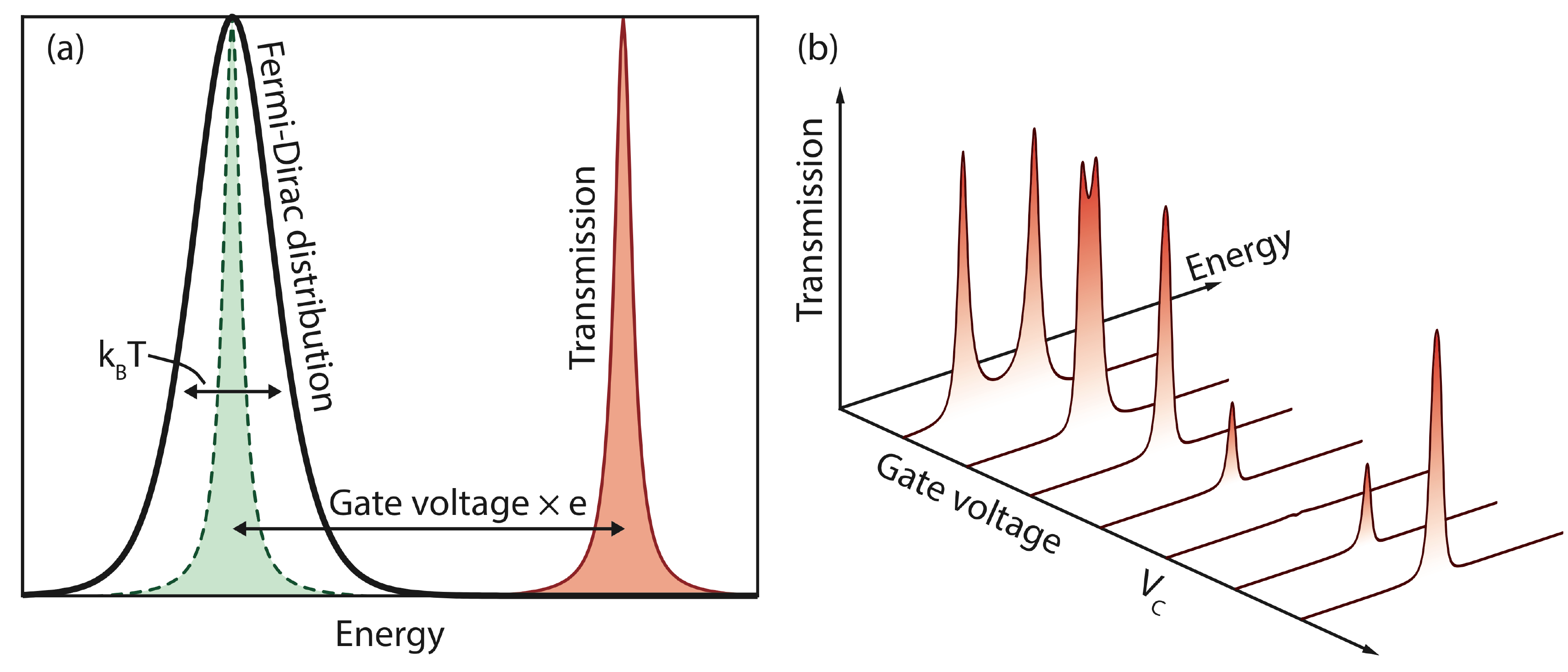}
\caption{
Two ways of switching the current with the gate. (a) The current is given by overlap between the transmission and the distribution function. The width of the distribution function $f_\text{L} - f_\text{R}$ is limited by the electron temperature. The current is switched on when the transmission (green curve) and the distribution function has a significant overlap; and the current is switched off when the transmission (red curve) is moved away from the distribution function. The gate voltage for switching the current is then limited by the temperature. (b) The gate varies the form of the transmission function, and at a specific gate voltage $V_\text{c}$ (or within a range of the gate voltage), the transmission function is greatly suppressed in (rather than moved from) the interval of the distribution function.
}
\label{fig:transmission}
\end{figure}

The current through a molecule can be expressed as $I = (2e/2\pi)\int_0^\infty dE T(E)[f_\text{L}(E) - f_\text{R}(E)]$~\cite{NazarovBook}. This is the integration of the transmission $T(E)$ and the Fermi-Dirac distribution $f_\alpha(E) = [ 1 + e^{(E-\mu_\alpha)/k_\text{B}T} ]^{-1}$, $\alpha = \text{L},\text{R}$ in the left and the right contact, where $\mu_\alpha$ is the electrochemical potential of the contact-$\alpha$, and $T$ is the electron temperature. While the distribution function is only determined by source and drain voltages (i.e.~$\mu_\text{L}$ and $\mu_\text{R}$) and the temperature $T$, the transmission can be affected by many factors, e.g.~ the internal dynamics of the molecule, the spectrum of contacts and the tunnel-coupling between the molecule and contacts. The current through the molecule can be controled by an electrostatic gate in two ways (see Fig.~\ref{fig:transmission}): i) the gate alters the energy of electrons in the molecule, i.e. shifting the transmission function along the energy axis, which changes the overlap between the transmission and the distribution function; or ii) the gate tunes the actual form of the transmission function, e.g.~the transmission peak is still in the distribution window but the height is suppressed. Both mechanisms influence the current as a function of the gate voltage. If the first mechanism dominates, the gate voltage for switching on and off the current is limited by the electron temperature, i.e.~the change of the gate voltage ($\times e$) must be larger than the width of the Fermi-Dirac distribution function of the leads. In order to seek a molecular transistor that is not limited by the Fermi-Dirac distribution of electrons in the contacts, we will therefore focus on the case in which the second mechanism plays a dominant role.

\section{Two-level molecule}

To investigate the second mechanism, we first consider a molecule with two localised single-electron states L and R~\cite{Markussen2010}, which are only coupled with the left and right contacts, respectively. The two localised states are coupled with each other via the tunnelling, hence, an electron can tunnel through the device from the left contact --- state L --- state R --- right contact. The Hamiltonian describing such a molecule is
$$H = \left( \begin{array}{cc}
E_\text{L} & J \\
J & E_\text{R}
\end{array} \right),$$
where $E_\text{L}$ and $E_\text{R}$ are respectively on-site energies and $J$ is the coupling strength. Here, we focus on the case where electron-electron interactions are negligible. Instead of varying the energy of all electrons in the molecule (i.e. shifting the total transmission function), we assume the gate only changes the energy of electrons in the state R, i.e.~$E_\text{R}$. In this case the amplitude of the transmission function will be tuned with the gate voltage. The transmission in this two-state model is
$$T(E) = \frac{J^2\gamma^2}{[(E-E_+)(E-E_-) + \gamma^2/4]^2 + \Delta E^2\gamma^2/4},$$
where $E_\pm = (E_\text{L}+E_\text{R})/2 \pm \Delta E /2$ are eigenenergies, $\Delta E =  \sqrt{(E_\text{L}-E_\text{R})^2 + 4J^2}$ is the difference between two eigenenergies, and $\gamma$ is the tunnel-coupling strength between the molecule and contacts (see Sec.~\ref{sec:twostate})~\cite{Meir1992, Thygesen2006, Lambert2015}. For simplicity, we only discuss the case $\gamma \leq 2J$, however, the conclusion is similar for the case $\gamma > 2J$. If $\gamma \leq 2J$, the maximum value of the transmission is $T_\text{max} = 4J^2/\Delta E^2$, which reaches $1$ when $E_\text{L} = E_\text{R}$ and decreases as $T_\text{max} \sim 4J^2/(eV_\text{g})^2$ when $eV_\text{g} \gg J$. Here we take the theoretical limit where $E_\text{L} - E_\text{R} \sim eV_\text{g}$ and $V_\text{g}$ is the gate voltage. In this limit, the gate voltage required for switching off the current is only determined by the coupling $J$ and independent with the electron temperature of two contacts. In a more realistic case, the detuning between the left and right localised state will only be a fraction of the applied gate voltage and a larger gate voltage will be required to switch the current.

\section{Four-level molecule}

\begin{figure}[tbp]
\centering
\includegraphics[width=1\linewidth]{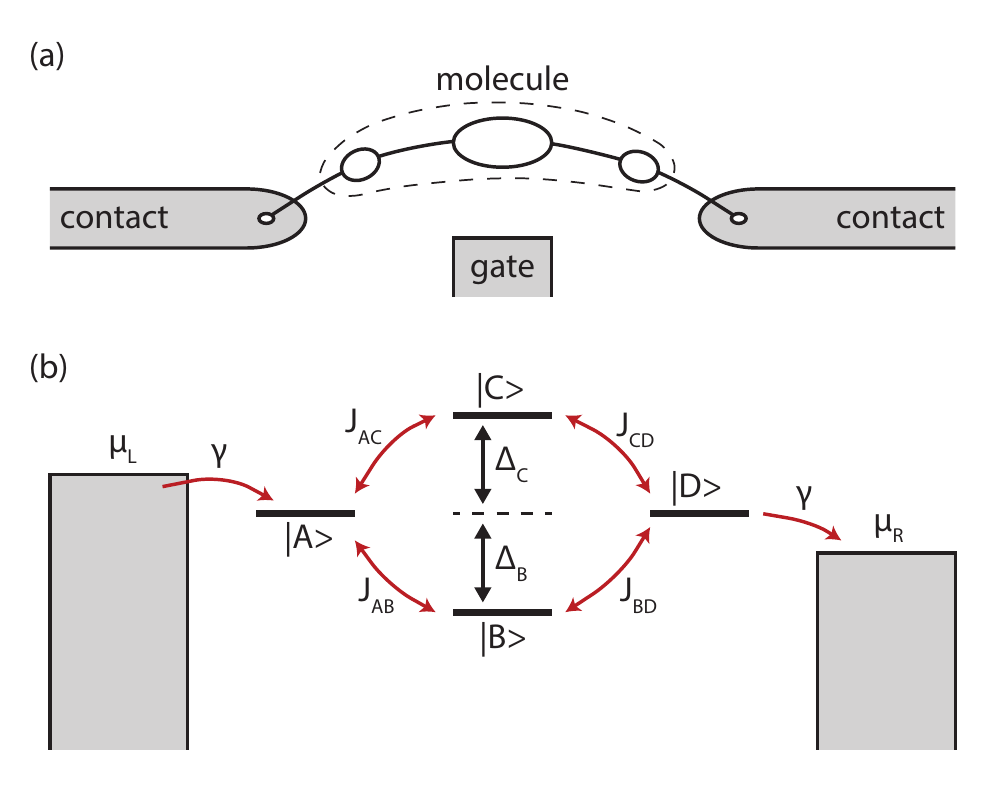}
\caption{
An interference-based molecular transistor and its level structure. States A and D are localised sates of side groups, and states B and C are HOMO and LUMO states of the centre group.
}
\label{fig:scheme}
\end{figure}

Interference allows for more efficient switching of the current. If the transmission vanishes due to complete destructive interference at a gate voltage $V_\text{c}$, any finite change of the gate voltage will result in the breaking of this interference condition and thus raise the current from zero to a finite intensity. Although the destructive-interference current will never completely vanish as a result of imperfections in the practical situation, we still expect the current to be strongly suppressed when the interference condition is met. In the following, we will show that destructive interference can exist in molecules with four localised states, and that the current is switched more efficiently than in a two-state molecule.

For interference to occur, there need to be at least two paths for electron transport. Here, we consider molecules with two side groups and a central group as shown in Fig.~\ref{fig:scheme}(a). Each side group is coupled with the corresponding contact, and there is no direct tunnelling between the central group and contacts. We assume only one single-electron state of each side group and the highest occupied molecular orbital (HOMO) and lowest unoccupied molecular orbital (LUMO) states of the central group are involved in the transport, while all other states are effectively irrelevant due to a large energy difference. HOMO and LUMO states of the central group are tunnel-coupled with the side-group states. The transport in this kind of molecule can be described by a model of four localised states as shown in Fig.~\ref{fig:scheme}(b). There are two pathways for electrons to tunnel through the molecule: Left --- HOMO --- Right and Left --- LUMO --- Right. When the energy of the HOMO (LUMO) state is lower (higher) than side-group states, an electron going through the HOMO (LUMO) path acquires a positive (negative) phase because the phase obtained in quantum evolution is $e^{-iEt/\hbar}$, where $E$ is the energy and $t$ is the evolution time. When the energy differences between side-group states and central-group states are much larger than the tunnel coupling, the phase difference between two paths is approximately $\pi$, and the destructive interference occurs.

The interference in the four-state model can also be understood using perturbation theory. When the energy differences $\Delta_\text{B}$ and $\Delta_\text{C}$ are much larger than the tunnel coupling matrix elements $J_{\beta,\beta'}$ [Fig.~\ref{fig:scheme}(b)], states B and C can be adiabatically eliminated, and the two side-group states A and D are effectively directly coupled with an effective tunnel coupling matrix element $J_\text{eff} = J_\text{AB}J_\text{BD}/\Delta_\text{B} - J_\text{AC}J_\text{CD}/\Delta_\text{C}$. Here, we have assumed that states A and D are nearly degenerate, i.e.~the energy difference between states A and D is much smaller than $\Delta_\text{B}$ and $\Delta_\text{C}$. We also need to assume that coupling strengths between side groups and contacts are much smaller than $\Delta_\text{B}$ and $\Delta_\text{C}$, so that on-site energies of side-group states are well defined. When $J_\text{eff} = 0$, the amplitudes of the two interference arms are identical, i.e.~the destructive interference is complete; according to our discussions of the two-state model, the transmission ($\propto J_\text{eff}^2$) is then zero, so the current is switched off.

The perturbation theory is only an approximation method, i.e.~the transmission is not exactly zero even when the condition $J_\text{eff} = 0$ is satisfied. For simplicity, in the following we assume states A and D are degenerate and all four tunnel coupling matrix elements are identical, i.e.~$J_{\beta,\beta'} = J$. In this case the condition for complete destructive interference is $\Delta_\text{B} = \Delta_\text{C}$. When this condition is satisfied, the transmission in the four-state model is
$$T(E) = \frac{4E^2J^4\gamma^2}{(E^2+\gamma^2/4)[(x^2-4J^2)^2E^2 + x^4\gamma^2/4]},$$
where $x^2 = E^2-\Delta^2$, $\gamma$ is the coupling strength for tunneling between the side groups and the contacts, $\Delta_\text{B} = \Delta_\text{C} = \Delta$, and without loss of generality we set the on-site energy of side-group states to zero (see Sec.~\ref{sec:fourstate}). We are interested in the case where the current results predominantly from transmission of electrons near to side-group energy levels ($E\sim 0$), i.e.~the distribution function $f_\text{L} - f_\text{R}$ is centered around side-group levels and its width is smaller than the HOMO-LUMO gap of the central group (both the bias voltage and the temperature are low enough compared with the gap). In this case, we can rewrite the transmission as
$$T(E) \simeq \frac{4E^2J^4\gamma^2}{(E^2+\gamma^2/4)^2\Delta^4},$$
and the maximum of the transmission around $E = 0$ is $T_\text{max} \simeq 4J^4/\Delta^4$.

To switch the current, an electrostatic gate tunes the energy levels in the four-state molecule. Similar to the two-state model, we assume the gate only changes the energy of electrons in the central group, i.e.~$\Delta_\text{B}$ and $\Delta_\text{C}$. Moving the HOMO and LUMO levels of the central group with the gate voltage will break the condition $\Delta_\text{B} = \Delta_\text{C}$, resulting in an increase of the current. When the LUMO (HOMO) level is aligned with side-group levels, the HOMO (LUMO) state is effectively decoupled from other states due to the large energy difference, and the transmission is mainly due to the LUMO (HOMO) state and two side-group states. Then the transmission is increased to
$$T(E) = \frac{J^4\gamma^2}{(E^2+\gamma^2/4)[(E^2-2J^2)^2 + E^2\gamma^2/4]},$$
which has the maximum value $T_\text{max} = 1$ (see Sec.~\ref{sec:threestate}). Therefore, by varying the gate voltage for $eV_\text{g} \sim \Delta$ (assuming that the shift of the energy levels is proportional to the gate voltage), the transmission can be switched for $\sim 4\log_{10}(J/\Delta)$ decades. Compared with the two-state model, in which the OFF current decreases quadratically with the gate voltage, the current in the four-state model is more sensitive to the gate voltage, and the OFF current scales as $\sim J^4/(eV_\text{g})^4$. We would like to note that, unlike in the two-state model, here the OFF current is only given by the gate voltage around a specific value $V_\text{c}$, and varying the voltage to either direction results in an increase of the current.

\begin{figure*}[tbp]
\centering
\includegraphics[width=0.8\linewidth]{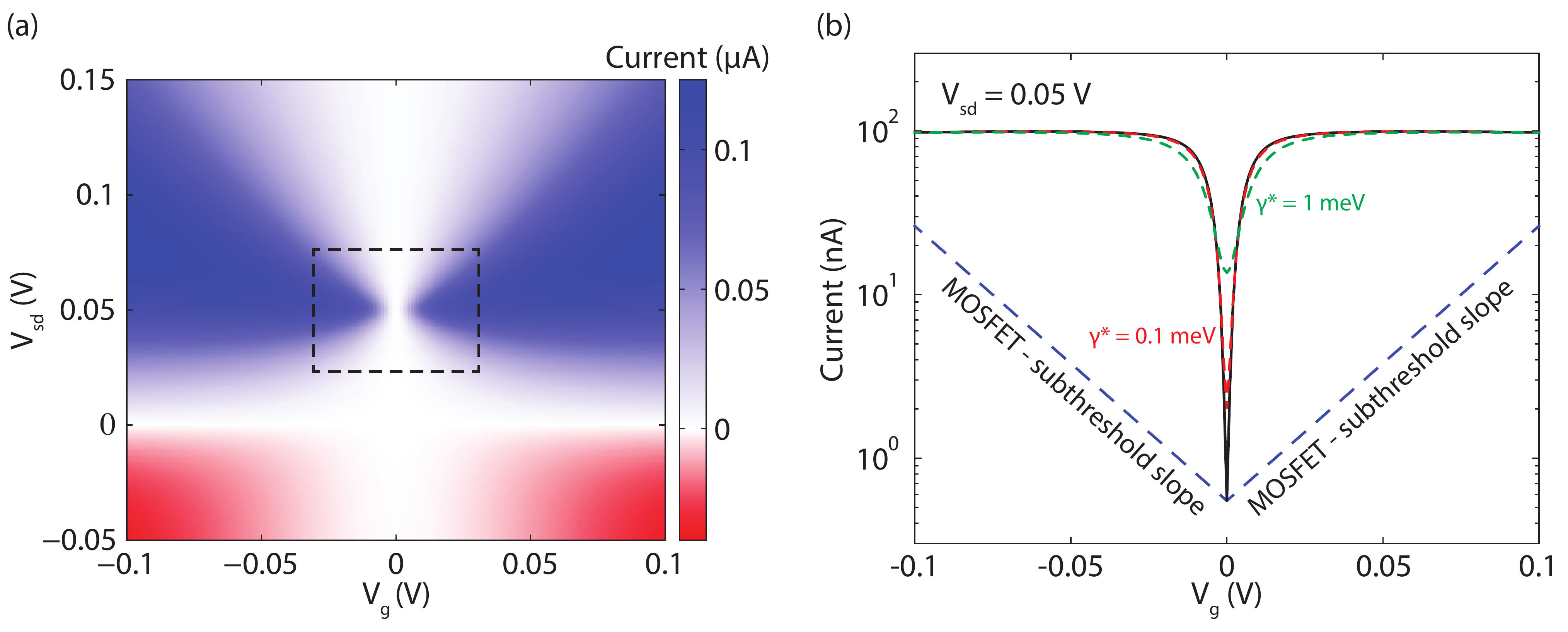}
\caption{
Current going through the interference-based molecular switch. The HOMO-LUMO gap of the centre group is $\Delta_\text{gap} = 1\text{ eV}$. Each side-group state is capacitively coupled with the corresponding contact with the lever arm $0.2$. When the basis voltage $V_\text{sd} = 0$, states A and D have an energy difference $\delta = 0.01\text{ eV}$, which vanishes at $V_\text{sd} = 0.05\text{ V}$. The basis voltage is anti-symmetrically applied to the molecule, i.e.~$\mu_\text{L} = eV_\text{sd}/2$ and $\mu_\text{R} = -eV_\text{sd}/2$. The tunnelling coupling between four localised states is $J = 0.1\text{ eV}$ and the coupling between side groups and contacts is $\gamma = 1\text{ meV}$. The temperature is $300\text{ K}$. We have assumed that $\Delta_B = \Delta_C$ when the gate voltage $V_\text{g} = 0$, i.e.~the gate voltage of maximum interference is $V_\text{c} = 0$. See Sec.~\ref{sec:current} for details of the model.
}
\label{fig:current}
\end{figure*}

The current for giving bias voltage and gate voltage in the four-state model is shown in Fig.~\ref{fig:current}. In our calculation, we have assumed that the HOMO-LUMO gap of the central group $\Delta_\text{gap} = \Delta_\text{B} + \Delta_\text{C}$ is a constant, and that the gate voltage varies central-group levels in parallel with a lever arm $1$, i.e.~$\Delta_\text{B} = -\Delta_\text{gap}/2 - e(V_\text{g}-V_\text{c})$ and $\Delta_\text{C} = \Delta_\text{gap}/2 - e(V_\text{g}-V_\text{c})$, and all other parameters, e.g.~$\gamma$ and $J$, are independent with the gate voltage. If the lever arm is not $1$, the current is less sensitive to the gate voltage, but the quartic scaling still applies. Side groups may be capacitively coupled to contacts, so two side-group states may only be degenerate if the basis voltage neutralizes the raw splitting between side-group states. Including capacitive coupling to contacts, the current shows a butterfly shape [Fig.~\ref{fig:current}(a)] as a function of gate and bias voltage, and the interference is maximised at the centre of the butterfly. Because of interference, the current is not a monotonic function of the basis voltage, i.e.~the conductance can be negative upon reaching the interference condition. We find that the current [black curve in Fig.~\ref{fig:current}(b)] can be much more sensitive to the gate voltage than in the case of a MOSFET; in fact, by changing the gate voltage by $\sim 20\text{ mV}$, the current is switched by two decades.

\section{Discussion}

In our previous discussion, we have neglected noise and interactions between electrons. The complete destructive interference is due to the exact cancellation of amplitudes of two arms, which relies on the appropriate energy difference between centre-group states and side-group states in the four-state model. Electric-field noise and couplings to phonon modes may broadening the line widths of these localised states, so that energies of these states are not well defined. As a result, the minimum current will be larger than expected. At the point of maximum interference, a small variance of energy levels $\gamma^*$ could cause the effective tunnelling between side-group states with the strength $\sim \gamma^*J^2/\Delta^2$. According to our analysis of the two-state model, the maximum transmission given by this effective tunnelling is $\sim \gamma^*J^4/\gamma \Delta^4$. Therefore, if the noise-induced broadening $\gamma^*$ is much smaller than the broadening induced by the coupling to contacts $\gamma$, the performance of the switch will not be dramatically affected by noise. In Fig.~\ref{fig:current}(b), we considered the effect of noise-induced broadening (see dashed red and green curves) using the master equation approach~\cite{Sun1999, BreuerBook}. We find that, with the broadening $\gamma^* = \gamma = 1\text{ meV}$, the current in the interference-based switch is still more sensitive to the gate voltage than in a MOSFET. When interactions between electrons are weak, they will simply results in broadening of the line widths. When interactions are strong and the repulsion energy is  larger than the tunnel coupling and the Fermi-Dirac distribution window, multi-electron transports is suppressed and electrons tunnel through the molecule one by one \cite{Bogani2008}.  In this case, the dynamic of the single electron is as the same as for interaction-free electrons, so that our analysis of the destructive interference is still valid.

Up to this point we have assumed that the energy levels shift with a lever arm $1$. While in reality this value is unachievable, values of 0.1 have been reported for single-molecule devices with a 3 nm thin gate oxide layer \cite{Osorio2007} and as much as 0.3 for co-planar graphene gate electrodes \cite{Puzckarski2015}, making our gating scheme feasible in a nanometre sized device geometry. Moreover, different electrostatic gate coupling for different molecular orbitals has been theoretically predicted \cite{Kaabjerg2008} and demonstrated experimentally \cite{Perrin2013}. In particular for a molecular design consisting of a central-group with weakly bound side-groups the electrostatic gate coupling to the central group is expected to be much larger than the electrostatic gate coupling to the side-groups.

The parameters which we have used for the results shown in Fig.~\ref{fig:current} are illustrative. Specific devices may perform worse or better. Nevertheless the calculations show how molecular quantum interference can yield a subthreshold slope that is not thermally limited and is almost independent of temperature. Huge advances are being made in the stability and reproducibility of single-molecule devices, enabled by the use of nanogaps in graphene ribbons for electrodes~\cite{Mol2015}. Conditions are therefore excellent to develop molecular electronics with the potential to reduce the energy consumption of switching for ICT. There is plenty of thermodynamic room for improvement, and quantum interference may enable some of this potential to be realised.

\section{Appendix}

\subsection{Two-level molecule}
\label{sec:twostate}

The transmission is given by~\cite{Thygesen2006}
\begin{eqnarray}
T(E) = \Tr [G^\text{r}(E)\Gamma_\text{L}(E)G^\text{a}(E)\Gamma_\text{R}(E)],
\label{eq:transmission}
\end{eqnarray}
where
\begin{eqnarray}
G^\text{r}(E) = \Tr [ES - H - \Sigma_\text{L}(E) - \Sigma_\text{R}(E)]^{-1}
\end{eqnarray}
is the retarded Green's function, $G^\text{a}(E) = G^\text{r}(E)^\dag$ is the advanced Green's function, and $\Gamma_\text{L,R}(E) = i[\Sigma_\text{L,R}(E) - \Sigma_\text{L,R}(E)^\dag]$ are matrices describing couplings to contacts. Here, $H$ is the single-particle Hamiltonian, $S$ is the overlap matrix, and $\Sigma_\text{L,R}(E)$ are self-energies. When single-particle states are orthonormal, $S = 1$. For simplification, we assume that self-energies are independent with the energy $E$, and the real part of self-energies has been included in the Hamiltonian.

For the two-state model, the Hamiltonian
\begin{eqnarray}
H = \left( \begin{array}{cc}
E_\text{L} & J \\
J & E_\text{R}
\end{array} \right),
\end{eqnarray}
and coupling matrices
\begin{eqnarray}
\Gamma_\text{L} = \left( \begin{array}{cc}
\gamma & 0 \\
0 & 0
\end{array} \right),
\end{eqnarray}
\begin{eqnarray}
\Gamma_\text{R} = \left( \begin{array}{cc}
0 & 0 \\
0 & \gamma
\end{array} \right).
\end{eqnarray}
Here, we have assumed that tunnelling couplings between the molecule and contacts are independent with the energy. Then, using Eq.~(\ref{eq:transmission}), we can obtain the transmission
\begin{eqnarray}
&& T(E) \notag \\
&=& \frac{J^2\gamma^2}{[(E-E_+)(E-E_-) + \gamma^2/4]^2 + \Delta E^2\gamma^2/4},
\end{eqnarray}
where $E_\pm = (E_\text{L}+E_\text{R})/2 \pm \Delta E /2$ and $\Delta E =  \sqrt{(E_\text{L}-E_\text{R})^2 + 4J^2}$.

When $\gamma \leq \Delta E$, the transmission is maximised at energies $E = (E_\text{L}+E_\text{R})/2 \pm \sqrt{\Delta E^2 - \gamma^2}/2$, and the maximum value of the transmission is $T_\text{max} = 4J^2/\Delta E^2$; when $\gamma > \Delta E$, the transmission is maximised at the energy $E = (E_\text{L}+E_\text{R})/2$, and the maximum value of the transmission is $T_\text{max} = 16J^2\gamma^2/(\gamma^2 + \Delta E^2)^2$.

If $\gamma \leq 2J$, the condition $\gamma \leq \Delta E$ is always satisfied, i.e.~$T_\text{max} = 4J^2/\Delta E^2$. In this case, $T_\text{max} = 1$ when $E_\text{L} - E_\text{R} = 0$; and
\begin{eqnarray}
T_\text{max} \simeq \frac{4J^2}{(E_\text{L} - E_\text{R})^2}
\label{eq:Tmax}
\end{eqnarray}
when $\abs{E_\text{L} - E_\text{R}} \gg J$.

If $\gamma > 2J$, the condition $\gamma \leq \Delta E$ is not satisfied when $E_\text{L} - E_\text{R} = 0$. In this case, the transmission cannot reach $1$. When $E_\text{L} - E_\text{R} = 0$, $T_\text{max} = 16J^2\gamma^2/(\gamma^2 + 4J^2)^2$. However, when $\abs{E_\text{L} - E_\text{R}} \gg \gamma$, the condition $\gamma \leq \Delta E$ is satisfied again, and the maximum transmission is given by Eq.~(\ref{eq:Tmax}). Therefore, Eq.~(\ref{eq:Tmax}) always describes the behaviour of the maximum transmission at large gate voltage.

\subsection{Four-level molecule}
\label{sec:fourstate}

The Hamiltonian of the four-state model is
\begin{eqnarray}
H = \left( \begin{array}{cccc}
E_\text{A} & J_\text{AB} & J_\text{AC} & 0 \\
J_\text{AB} & E_\text{B} & 0 & J_\text{BD} \\
J_\text{AC} & 0 & E_\text{C} & J_\text{CD} \\
0 & J_\text{BD} & J_\text{CD} & E_\text{D}
\end{array} \right),
\end{eqnarray}
and coupling matrices are
\begin{eqnarray}
\Gamma_\text{L} = \left( \begin{array}{cccc}
\gamma & 0 & 0 & 0 \\
0 & 0 & 0 & 0 \\
0 & 0 & 0 & 0 \\
0 & 0 & 0 & 0
\end{array} \right),
\end{eqnarray}
\begin{eqnarray}
\Gamma_\text{R} = \left( \begin{array}{cccc}
0 & 0 & 0 & 0 \\
0 & 0 & 0 & 0 \\
0 & 0 & 0 & 0 \\
0 & 0 & 0 & \gamma
\end{array} \right).
\end{eqnarray}
Taking $E_\text{A} = E_\text{D} = 0$, $E_\text{B} = -\Delta_\text{B} = -\Delta$, $E_\text{C} = \Delta_\text{C} = \Delta$ and $J_\text{AB} = J_\text{AC} = J_\text{BD} = J_\text{CD} = J$, and using Eq.~(\ref{eq:transmission}), we can obtain the transmission
\begin{eqnarray}
&& T(E) \notag \\
&=& \frac{4E^2J^4\gamma^2}{(E^2+\gamma^2/4)[(x^2-4J^2)^2E^2 + x^4\gamma^2/4]},
\end{eqnarray}
where $x^2 = E^2-\Delta^2$.

When $\Delta \gg E,J$, we have $x \gg J$. Then, the Taylor expansion gives
\begin{eqnarray}
T(E) &=& \frac{4E^2J^4\gamma^2}{(E^2+\gamma^2/4)^2x^4}
\left( 1 + \frac{8E^2}{E^2+\gamma^2/4}\frac{J^2}{x^2} + \cdots \right) \notag \\
&=& \frac{4E^2J^4\gamma^2}{(E^2+\gamma^2/4)^2\Delta^4}
\left( 1 + 2\frac{E^2}{\Delta^2} + \cdots \right) \notag \\
&& \times \left( 1 + \frac{8E^2}{E^2+\gamma^2/4}\frac{J^2}{x^2} + \cdots \right).
\end{eqnarray}
Neglecting small terms, we get
\begin{eqnarray}
T(E) &\simeq & \frac{4E^2J^4\gamma^2}{(E^2+\gamma^2/4)^2\Delta^4},
\end{eqnarray}
which is maximised at $E = \gamma/2$, and the maximum value is $T_\text{max} \simeq 4J^4/\Delta^4$.

\subsection{Three-level molecule}
\label{sec:threestate}

When one of states B and C in the four-state model is ignored, and all other three states have the same on-site energy, the Hamiltonian reads
\begin{eqnarray}
H = \left( \begin{array}{ccc}
0 & J & 0 \\
J & 0 & J \\
0 & J & 0
\end{array} \right),
\end{eqnarray}
and coupling matrices read
\begin{eqnarray}
\Gamma_\text{L} = \left( \begin{array}{cccc}
\gamma & 0 & 0 \\
0 & 0 & 0 \\
0 & 0 & 0
\end{array} \right),
\end{eqnarray}
\begin{eqnarray}
\Gamma_\text{R} = \left( \begin{array}{cccc}
0 & 0 & 0 \\
0 & 0 & 0 \\
0 & 0 & \gamma
\end{array} \right).
\end{eqnarray}
Then, using Eq.~(\ref{eq:transmission}), we obtain the transmission
\begin{eqnarray}
T(E) = \frac{J^4\gamma^2}{(E^2+\gamma^2/4)[(E^2-2J^2)^2 + E^2\gamma^2/4]},
\end{eqnarray}
is maximised at $E = 0$ with the maximum value $T_\text{max} = 1$.

\subsection{Current in the four-level molecule}
\label{sec:current}

To obtain Fig.~\ref{fig:current}, we have taken $E_\text{A} = - \delta/2 + l_\text{S}eV_\text{sd}/2$, $E_\text{D} = \delta/2 - l_\text{S}eV_\text{sd}/2$, $E_\text{B} = - \Delta_\text{gap}/2 - l_\text{C}eV_\text{g}$ and $E_\text{C} = \Delta_\text{gap}/2 - l_\text{C}eV_\text{g}$ in the four-state model. Here, lever arms $l_\text{S} = 0.2$ and $l_\text{C} = 1$. There are two different approaches of calculating the current. The current without the broadening of levels [Fig.~\ref{fig:current}(a) and the solid black curve in Fig.~\ref{fig:current}(b)] is calculated using the formula $I = (2e/2\pi)\int_0^\infty dE T(E)[f_\text{L}(E) - f_\text{R}(E)]$, where the transmission $T(E)$ is given by Eq.~(\ref{eq:transmission}). The current with the broadening of levels [dashed red and green curves in Fig.~\ref{fig:current}(b)] is calculated using the master equation approach~\cite{Sun1999}. The master equation also can be used to calculate the current without the broadening of levels, in which case the difference between results given by two approaches is negligible.

The master equation reads
\begin{eqnarray}
\partial _t \rho = -\frac{i}{\hbar}[H,\rho] + \mathcal{T}\rho + \mathcal{L}\rho,
\label{eq:ME}
\end{eqnarray}
where $\mathcal{T}$ describes the tunnelling of electrons between the molecule and contacts, and $\mathcal{L}$ describes the broadening of four localised single-electron levels. The Tunnelling term has two parts, i.e.~$\mathcal{T} = \mathcal{T}_\text{L} + \mathcal{T}_\text{R}$, which respectively correspond to couplings to left and right contacts and can be expressed as
\begin{eqnarray}
\mathcal{T}_\alpha \rho &=& \frac{1}{2\hbar}\sum_{m,n} \gamma_{\alpha;m,n}
\{f_\alpha(E_n) (a_n^\dag \rho a_m - a_m a_n^\dag \rho) \notag \\
&&+ f_\alpha(E_m) (a_n^\dag \rho a_m - \rho a_m a_n^\dag) \notag \\
&&+ [1-f_\alpha(E_n)](a_m \rho a_n^\dag - \rho a_n^\dag a_m) \notag \\
&&+ [1-f_\alpha(E_m)] (a_m \rho a_n^\dag - a_n^\dag a_m \rho) \}.
\end{eqnarray}
Here, $\alpha = \text{L},\text{R}$ denotes two contacts; $\{a_m\}$ are annihilation operators of electrons in eigenstates (with the eigenenergies $\{E_m\}$) of the single-electron Hamiltonian $H$, $a_m = \sum_{\beta}u_{m,\beta}^* a_\beta$, $\beta  = \text{A},\text{B},\text{C},\text{D}$, and $\{a_\beta\}$ are annihilation operators of electrons in four localised states. This tunnelling term correspond to the second-order approximation of the Nakajima-Zwanzig equation~\cite{BreuerBook}, and parameters $\{\gamma_{\alpha;m,n}\}$ are given by
\begin{eqnarray}
\gamma_{\text{L};m,n} &=& \gamma u_{m,\text{A}} u_{n,\text{A}}^*, \\
\gamma_{\text{R};m,n} &=& \gamma u_{m,\text{D}} u_{n,\text{D}}^*.
\end{eqnarray}
We model the broadening of levels as pure dephasing, i.e.~
\begin{eqnarray}
\mathcal{L} \rho &=& \frac{\gamma^*}{4\hbar}\sum_\beta
[(1-2a_\beta^\dag a_\beta) \rho (1-2a_\beta^\dag a_\beta) - \rho].
\end{eqnarray}
The current (from right to left) is then given by
\begin{eqnarray}
I &=& \frac{2e\gamma}{\hbar} [\sum_{m} \abs{u_{m,\text{A}}}^2 f_\text{L}(E_m) - n_\text{A} ],
\end{eqnarray}
where $n_\text{A} = \mean{a_\text{A}^\dag a_\text{A}}$ is the average occupation of the single-electron state A in the steady state of Eq.~(\ref{eq:ME}).

\begin{acknowledgments}
This work was supported by the EPSRC Platform Grant `Molecular Quantum Devices' (EP/J015067/1), the EPSRC National Quantum Technology Hub in Networked Quantum Information Processing, and Templeton World Charity Foundation. The opinions expressed in this publication are those of the authors and do not necessarily reflect the views of Templeton World Charity Foundation. The authors would like to thank Prof. Gerard Milburn and Prof. Colin Lambert for useful discussions. 
\end{acknowledgments}

\end{document}